\documentclass[aps,prb,preprint,preprintnumbers,amsmath,amssymb,floats,a4paper, margin=2.5cm]{revtex4}
\usepackage{graphicx}% Include figure files
\usepackage{dcolumn}% Align table columns on decimal point
\usepackage{bm,epstopdf,soul}
\usepackage{amsmath}
\usepackage{color}
\usepackage{soul}

\usepackage{times}

\usepackage{amsmath}
\usepackage{color}
\usepackage{soul}

\def\baselinestretch{1.5}
\usepackage{fancyhdr}

\fancypagestyle{fancy}{% normal pages
\fancyhf{}
\renewcommand{\headrulewidth}{0pt}
\fancyhead[R]{\thepage}
\renewcommand{\headrulewidth}{0pt}
}

\begin{document}

%\preprint{}

\title{Hydrogen induces chiral conduction channels in the topological magnet} %72 characters +  8sp

\author{Afrin N. Tamanna$^1$, Ayesha Lakra$^1$, Xiaxin Ding$^1$, Entela Buzi$^1$, Kyungwha Park$^2$, Kamil Sobczak$^{3}$, Haiming Deng$^1$, Gargee Sharma$^4$, Sumanta Tewari$^5$
\& Lia Krusin-Elbaum$^{1,6,{\dagger}}$}

\vspace{25mm}

\affiliation{$^1$Department of Physics, The City College of New York - CUNY, New York, New York 10031, USA}
\affiliation{$^2$Department of Physics, Virginia Tech, Blacksburg, Virginia 24061, USA}
\affiliation{$^3$Faculty of Chemistry, University of Warsaw, 02-089 Warsaw, Poland}
\affiliation{$^4$School of Physical Sciences, Indian Institute of Technology–Mandi, Himachal Pradesh 175005, India}
\affiliation{$^5$Department of Physics and Astronomy, Clemson University, Clemson, South Carolina 29634, USA}
\affiliation{$^6$City University of New York Graduate Center, New York, New York 10016, USA}

\begin{abstract}
\vspace{5mm}
\noindent \small \textbf{{{Chirality, a characteristic handedness that distinguishes `left’ from ‘right’, cuts widely across all of nature \cite{Chiral2-1990},
from the structure of DNA \cite{chiral-bio2001} to opposite chirality of particles and antiparticles \cite{chiral-antimatter2003}.
In condensed matter chiral fermions have been
identified in Weyl semimetals \cite{Weyl-review2018} through their unconventional electrodynamics arising from `axial' charge imbalance between chiral Weyl nodes of topologically nontrivial electronic bands.
Up to now it has been challenging or impossible to create transport channels
of Weyl fermions in a single material that could be easily configured
for advancing chiral logic or spintronics \cite{Chir-DWlogic-Omari2019,Chir-spin_Parkin2021}.
Here we generate chirality-directed conduction channels in inversion-symmetric Weyl ferromagnet (FM) MnSb$_2$Te$_4$, emergent
from a deep connection between chirality in reciprocal and real space. % chiral transport.
We alter the bandstructure on-demand
with an intake and a subsequent release of ionic hydrogen ($\textrm{H}^{+}$) --- a process
we show to induce the tilt and rotation of Weyl bands.
The transformed Weyl FM states feature a doubled Curie temperature $\gtrsim 50$~K and an enhanced angular transport chirality synchronous with a rare field-antisymmetric longitudinal resistance
--- a low-field tunable `chiral switch' that roots in the interplay of Berry curvature \cite{BerryPhaseReview2010}, chiral anomaly \cite{PHE-Burkov2017} and
hydrogen-engendered mutation of Weyl nodes.}}}
\end{abstract}

%\keywords{}

\maketitle
\normalsize

Chirality is a fundamental property of quantum particles under broken symmetry intimately connected to their spins.
Chiral Weyl fermions ---  spin 1/2 massless particles introduced in relativistic field theories \cite{Weyl1929} but never found as elementary particles in nature --- have been predicted to
emerge as the low-energy excitations in condensed matter \cite{Ninomiya1983}.
Discovered only recently \cite{Hasan2015}, the Weyl quasiparticles have since been identified in
the so-named Weyl semimetals \cite{Weyl-review2018,Weyl-review2021},
a three-dimensional (3D) gapless topologically nontrivial phases with protected linear two-band crossings
in the Brillouin zone of the reciprocal space.

When either time reversal ($\mathcal{T}$) or inversion ($\mathcal{I}$) symmetries are broken, the crossing points (Weyl nodes) are doubly degenerate and come in pairs with node partners {having an opposite chirality % $\chi$,
set by  the parallel or antiparallel locking of quasiparticle spin to its momentum.}
The Weyl nodes are topologically robust (the two nodes
cannot be trivially gapped out) because they carry a topological charge, a quantized Berry flux through any surface enclosing them in momentum space \cite{Weyl-review2018}.
This charge, an integer topological invariant $C = \pm1$ called the Chern number, gives the electronic wavefunction a handedness.
Weyl nodes act as a sink or source of the Berry curvature \cite{BerryPhaseReview2010}, which in turn acts as a magnetic field in the momentum space.
All this has a huge impact on the electrodynamics of Weyl fermions, which has no counterpart in conventional metals
with phenomena such as the chiral anomaly --- the nonconservation of the chiral charge deriving from
a right-left imbalance between the occupancies $n^{3D}_{R/L}$ of the gapless, zeroth Landau levels with
opposite chiralities \cite{Weyl-review2018} under parallel electric and magnetic fields, namely $\frac {dn^{3D}_{R/L} }{dt} =  \pm{\frac{e^2}{h^2}}{\textbf{\textit{E}} \cdot \textbf{\textit{B}}}$ relevant to transport in 3D.

Although various Weyl and Weyl-like phases have been widely reported,
including Type II Weyl semimetals \cite{WEYL-TypeII2015,Weyl-TypeII2016}, hybrid and higher-order semimetals \cite{hybrid-Weyl2016,higher-orderWeyl2020},
and magnetic Weyl semimetals \cite{mKagome2019,FermiArcs2019},
current materials classification schemes \cite{TypeII-Bernevig2016} are yet to provide
a clear guiding principle or analytical framework for dialing in a specific Weyl fermion dynamics. This is particularly challenging in
magnetic ($\mathcal{T}$-symmetry broken) systems where {spin textures, by exerting ``axial gauge fields" \cite{axial2020} on Weyl fermions, may reorganize their collective behaviors} \cite{mWeylsAraki2020} in unexpected ways.

{In this work we report the discovery of {hydrogen-induced transformations }% transitions}
amongst unalike topological phases and spin textures. {The transfigured Weyl semimetal states are realized}
by tapping into the existing disorder landscape in the ferromagnetic MnSb$_2$Te$_4$ using the post-growth hydrogen intake-and-release technique we have recently
developed \cite{HCl-Haiming2022}, which turns Type I Weyl semimetal \cite{Weyl-review2018} with a moderately tilted Weyl nodes
into a {more tilted and rotated} Weyl-node system.}
{The modified Weyl FM state has {a twice-as-high} Curie temperature and displays strong field antisymmetry of the chiral-anomaly
driven transport \cite{PHE-Burkov2017} where enhanced angular chirality under in-plane magnetic fields translates
into out-of-plane field-antisymmetric chiral currents \cite{tilted_Weyls2019,AntisymMR-Weyl2021}, massively amplified at low fields}.
The role of ionic hydrogen is two-fold: {{through hybridization with Te} it alters magnetic anisotropy} and, by healing the ever-present dangling Mn-Te bonds,
enforces robust and reproducible transport asymmetries {driven by Berry curvature \cite{BerryPhaseReview2010,Berry-antisym2022}}.
The latter puts into a sharp relief the key role of bond/site disorder in controlling topological phases and shows
a path to a desired Weyl bandstructure %we may no longer be limited to a narrow set of natural materials, but
by the defect-altering of available materials with intrinsic phase space chirality.

A superabundance of exotic magnetic topological quantum states was recently predicted \cite{MBT-family2019} in an important XP$_2$C$_4$
class of $\mathcal{I}$-symmetric topological magnets (X = transition-metal or rare-earth element, P = group 15 pnictogens Bi or Sb, C = group 6A chalcogens Te, Se, or S), which now include a newly discovered
intrinsic axion \cite{axionMBT2020} and quantum anomalous Hall (QAH) insulators \cite{QAH-Haiming2020,QAH-MBT-Science2020} in MnBi$_2$Te$_4$, and a Type II
Weyl semimetal with one pair of Weyl nodes in antiferromagnetic (AFM) Mn(Bi$_{1-x}$Sb$_x$)$_2$Te$_4$
when Bi is substituted by Sb \cite{MBST-phase_diagram2019,AFM-MST-typeIIWeyl2021,MST-Weyl_Murakami2019}.
In XP$_2$C$_4$ magnetic X elements ideally are all arranged as a %n atomic
2D ferromagnetic sheet in a layered van der Waals (vdW) crystal structure consisting of  C-P-C-X-C-P-C septuple layers (SLs),
separated by vdW gaps yet coupled antiferromagnetically  \cite{Otrokov2019}.
In Mn(Bi$_{1-x}$Sb$_x$)$_2$Te$_4$, owing to the comparable sizes of Mn and Sb it is easy to form Sb$_{\textrm{Mn}}$ antisite (interchange) defects \cite{MST-site_mixing-Liu2021,Wimmer2021},
which can affect the interaction between SLs and turn the long-range order from AFM to FM.

Here we `redesign' a metallic-like nominally MnSb$_2$Te$_4$ (MST) single crystal with the uniform septuple layer (SL)
crystal structure {(Extended Data Figs. 1,2)}, grown as $p$-type (hole) conductor {(Fig. S1(A,B), Extended Data Fig. 3)}
typical of Sb-based systems \cite{MBST-phase_diagram2019}.
It contains ample antisite disorder {((Extended Data Fig. S1c)} and is slightly off nominal stoichiometry
({see Methods}), as requisite for the FM order  \cite{MST-site_mixing-Liu2021}.
The mean-field-like net magnetization $M(T)$,
set here by a joint action of SLs and the antisite-controlled inter-SL coupling,
aligns out-of-plane and onsets at the Curie temperature $T_C \cong 26~K$, in
correspondence with the characteristic cusp in the
temperature dependence of longitudinal resistance $R_{xx}(T)$ {(Fig. 1a)}, all consistent with prior reports \cite{MST-site_mixing-Liu2021}.

Our first immediately notable and surprising result is that insertion of ionic hydrogen {(Fig. 1b)} converts the as-grown FM MST into an antiferromagnet,
which is seen in the typical AFM signatures in magnetization
$M(T)$ and $R_{xx}(T)$  {(Fig. 1c)} as well as in the transverse (Hall) signal $R_{yx}(T)$ which tracks the peak in $M(T)$ at the Ne$\acute {\textrm{e}}$l temperature
$T_N \approx 20~\textrm{K}$, comparable to as-grown AFM MST {(Fig. 1d)}.  Here, the level of hydrogenation was controlled by timing the diffusion of
H$^+$ ions from a dilute (0.5 molar) HCl + H$_2$O = H$^\textrm{+}$(H$_2$O) + Cl$^\textrm{-}$
solution maintained at room temperature in which the samples were immersed for a period of minutes to hours (Methods) --- a process invented
and used by us previously \cite{HCl-Haiming2022} to reach charge neutrality and topological surface states through in-diffusion of H$^+$ into the bulk.
In MST, hydrogen intake also reduces carrier (hole) density in the AFM state {(Extended Data Figs. 3,4 and Supplementary Section I)} by donating electrons to the system, although the process appears self-limiting and
the MST system remains $p$-type throughout.

A reverse process --- the release of hydrogen {(Fig. 1e)} by a low-temperature annealing protocol \cite{HCl-Haiming2022} --- returns the system to an FM state,
however this new FM (FM 2)
is distinguished from the as-grown one (FM 1) in two spectacular ways: (i) the $T_C$ can be nearly doubled {up to} $\simeq 54~K$ (Fig. 1f and {Extended Data Fig. 4}) and (ii) the longitudinal resistance %$R_{xx}(H)$
turns from field-symmetric, i.e. $R_{xx}(H) = R_{xx}(-H)$ in FM 1 to field-antisymmetric $R_{xx}(H) = -R_{xx}(-H)$ in FM 2 (Figs. 1g,1i).
We remark that annealing of as-grown MST leaves $T_C$ and the field-symmetric $R_{xx}(H)$ intact {(Fig. S2)}.

To unpack the origins of this transformation, {cued by our magnetization measurements {(Fig. S3)}} we begin with asking whether (and how) the hydrogen intake-release process would affect magnetic anisotropy.
From X-ray photoelectron spectroscopy {(see Methods)} we determine that hydrogen mainly binds to Mn on the vacant Sb sites (Sb$_{\textrm{Mn}}$ antisites), and that after hydrogenation
cycles the broken Mn-Te bonds (abundant in as-grown MST) are passivated {(Extended Data Fig. 5, Fig. S4)}.  Our \textit{ab initio} (DFT) calculations confirm the formation of
Mn-H-Te moieties {(Fig. S5, Extended Data Fig. 6, Table S1)} in the FM 2 state, {which leads to doubling of $T_C$}
and indicates that {tilting magnetic easy axis towards normal to the Mn-H bonding direction lowers the energy of the ground state {(Supplementary Section I, Table S2)}.}

We test this further by first examining angular dependencies of the Hall signals under tilt $\theta$ of magnetic field {(see cartoon in Fig. 2)} in FM 1 and FM 2 {(Figs. 2a,2b)}.
In FM 1 the preferred magnetic alignment is along the crystalline $c$-axis, while in FM 2 the $45^\circ$ incline of the preferred direction is evident, albeit notably it appears only at low fields
in the magnetization reversal regime where magnetic domains proliferate. It suggests that the average domain wall direction
is impacted by the hydrogen-modified easy axis {(Figs. 2c,2d).} This is consistent with the longitudinal and Hall signals in FM 1 {(Figs. 2e,2f,{S6})} and FM 2 {(Figs. 2g,2h,{S7})}
that clearly show the $c$-axis alignment in FM 1 and a remarkable asymmetry at low fields in FM 2 --- a split in the $R_{xx}(\theta)$ into a dip (\emph{d}) at $\theta = 45^\circ$
and a peak (\emph{p}) slightly above $\theta = 90^\circ$ when magnetic field is grazing the $ab$-plane.
This asymmetry is seconded in $R_{yx}(\theta)$ below 50 mT as a kink-like shift to {a distinct $45^\circ$ angle}.
We point out that at higher fields the $c$-axis alignment in transport is entirely restored {(Figs. 2h,{S7})}.

For the field fully tilted to in-plane, a huge field-antisymmetric transverse signal $R_{yx}(H)$ dubbed
Planar Hall effect or PHE \cite{PHE-Burkov2017} is detected in both FM 1 and FM 2 {(Figs. 2i,2j}, also {{Extended Data Figs. 7(A,B), Fig. S8}}).
Moreover, an unusual in-plane anomalous $R_{yx}(H)$ loop (ARL) centered at $H = 0$ appears in both, tiny in FM 1
but is super-pronounced in FM 2. {Such ARL loops with $H \parallel ab$ are rare; it is a linear effect
(distinct from PHE) that emerges from an out-of-plane Berry curvature \cite{in-plane-AHE-Nature2022}
when the net magnetization is off the symmetry axis. {Notably, in FM 2 the magnetic easy axis is in the plane normal to the Mn-H bond direction {(Table S2)}.}}
Both the PHE and the ARL disappear at the respective $T_C$s of FM 1 and FM 2 {(Figs. 2i,2j)}. We point out that in FM1, where two field-symmetric peaks (\emph{p-p}) are seen in $R_{xx}(H\parallel c)$
the minuscule ARL width $H_c^{ab}$ vanishes at $T_C$. In FM 2 on the other hand, $H_c^{ab}$ is by comparison immense in the $T$-range where $R_{xx}(H\parallel c)$ is $H$-antisymmetric,
i.e. featuring a peak and a dip (\emph{p-d}) seen in {Fig. 1i}. {In FM 2, $H_c^{ab}$ becomes nearly null at $T_{dp} \simeq 40~\textrm{K}$ before reentering at the level of FM 1 {(Fig. 2k)},
where the symmetric \emph{p-p} behavior is also recovered.} {The resistance difference between the peak and the dip,
$\Delta^{p-d} R_{xx}(T)$ {(see Fig. 2l)} which switches to symmetric \emph{p-p} at $T_{dp}$ {(also Extended Data Fig. 7(A,B))}, confirms the correlation of both PHE and ARL with \emph{p-d}.}

The most striking visual of the connection between the peak and dip and the antisymmetry of PHE is obtained from its evolution with the tilt angle $\theta$. The \emph{p-d} field coordinates {(Fig. 2m)} are
very near zero field at first,  but then split at  the aforementioned angle $\theta = 45^\circ$ and
find a terminus at the antisymmetric `wings' of PHE when $\theta = 90^\circ$ ($H\parallel\textrm{ab}$), see {Figs. 2m,2n}.
This connection is reproducible and robust; it is also found after a repeat hydrogenation cycle stabilizing the hydrogen-transformed FM state {({Fig. S8}}).

From the above, we surmise that the in-plane transport holds the key to hydrogen-induced $H$-antisymmetric $R_{xx}(H)$ in MST for $H\parallel c$.
In service of this idea we next examine and compare transport parameters under in-plane ($\varphi$-angle) field rotation (see cartoon in {Fig. 3)} in the FM 1 and FM 2 states.
We immediately note the salient distinctions in the in-plane field dependencies: in FM1 the longitudinal magnetoconductance (MC) $\sigma_{xx}(H)$ is $H$-parabolic ($\propto  H^2$), while it is strictly linear in FM 2, excluding the low-field
(PHE-related) hysteretic region {(Figs. 3a,3b)}. {In theory \cite{Tewari_PRL2017,tilted_Weyls2019}
these MC behaviors have been associated with tilted Type I and Type II Weyl nodes respectively.} {We find that such field dependencies are also obtained by tuning the
internode/intranode scattering strength {(Supplementary Section II, Fig. S1(A,B)).}}
Here, the difference is also striking at low ($\leq$ 250 mT) fields, displaying symmetric MC in FM 1 and antisymmetric one in FM 2, {the latter likely reflecting the impact of tilting
that would allow the contributions from both, PHE and ARL.} {The dissimilarity is further notable in the 3D plots of $\sigma_{yx}(H,\varphi)$ --- {in contrast to FM 1 (Fig. 3c),
in FM 2 $\sigma_{yx}$ changes polarity at $\varphi = 90^\circ$ {(Fig. 3d)}.}

Our recorded $\varphi$ variations of the in-plane transverse and longitudinal conductances, $\sigma_{yx}(\varphi)  =  \Delta \sigma \sin (\varphi + \varphi_0)$
and $\Delta\sigma^{avg}_{xx}(\varphi) = \sigma_{\perp} + \Delta\sigma cos^2 (\varphi + \varphi_0)$, {are as anticipated in a Weyl system \cite{Tewari_PRL2017}}; they globally
match up in FM 1 {(Fig. 3e) and FM 2 (Fig. 3g)}, {albeit in FM 2 $\sigma_{yx}$ is phase-shifted by $90^\circ$ relative to FM 1 {(Supplementary Section II).}}
Here $\sigma_{yx}$ is the planar Hall signal (PHE) and $\Delta \sigma = \mid{\sigma_\perp - \sigma_\parallel}\mid$ is the anisotropy in conductivity due to chiral anomaly arising from the $\textbf{\textit{E}} \cdot \textbf{\textit{B}}$ term,
and thus {expected to be maximal for $\sigma_\parallel = \sigma_{xx} (\varphi = 0^\circ)$ and minimal for $\sigma_\perp = \sigma_{xx} (\varphi = 90^\circ)$.}}
{We note that PHE is a more reliable measure of chiral anomaly than a positive MC in {Figs. 3a and 3b,} which crucially depends on disorder modification of the inter- and intranode scattering events \cite{Tewari_PRB2023}}.
In what follows we show that these distortions are associated with the tilt mutation of Weyl nodes.

The ratio $\mathcal{R} =\frac{\Delta\sigma_{yx}(\varphi)}{\Delta\sigma^{avg}_{xx}(\varphi)}$ is exceptionally large in FM 1 {(Fig. 3f)} and truly giant in FM 2 {(Fig. 3h)}, as was predicted for PHE in a Weyl semimetal \cite{PHE-Burkov2017}.
We note that in both FM phases $\mathcal{R}$ is strongly field-dependent --- it nearly `blows up' at low fields in both.
There are, however, significant differences. A closer look at the $\sigma_{yx}(\varphi)$  in FM 2 reveals deviations from a simple $\cos \varphi$, evident in the Fourier transform {(Fig. 3i)} as two
frequencies {({see analysis in Fig. S9})} that are likely responsible for the distortion also manifest in $\sigma_{xx}(\varphi)$.

Generally, since the Lorentz invariance in condensed matter can be broken \cite{Weyl-review2018}, Weyl nodes are permitted to tilt. In  $\mathcal{T}$-symmetry-broken (magnetic) Weyl semimetals
Weyl node partners with different chiralities tilt in the opposite directions, producing currents that are linear and antisymmetric functions of magnetic field \cite{Tewari_PRL2017,tilted_Weyls2019}.
To get a sense of  the tilt directions we take an in-depth look at the angular variations of the in-plane transport coefficients {(Fig. 4)}.
In FM 1, polar plots of $\sigma^{avg}_{xx}(\varphi)$  {(Figs. 4b,4c)} and $\sigma_{yx}(\varphi)$ for the positive and negative values of $H$ {(Figs. 4d,4e)}
all have a `figure-eight' shape with the elongated direction in $\sigma^{avg}_{xx}$
rotated between high (4 T)  and low (0.5 T) values of $H$ from $\varphi = 0^\circ$ to $120^\circ$  but remaining slightly off $\varphi = 90^\circ$ in $\sigma_{yx}$ for all fields.
These features are captured with good fidelity in a semiclassical theory of {linear} transport in Weyl semimetals
when the ratio of internode/intranode scattering strength $\alpha$ is allowed to vary {(Supplementary Section II, Figs. S10-S11).}
From theory, we deduce the Weyl tilt \textbf{\textit{\^{t}}} in FM 1 to be of a moderately tilted Type I kind, nearly aligned with $x$-axis in the $xy$-plane {(Fig. 4a)}. The
parameter $\alpha$ here is nearly doubled from 1.5 at high $H$ to 2.65 at a lower field, consistent with intranode conductivity decreasing with increase in the magnetic field \cite{Tewari_PRB2023}.

The reorientation and modification of Weyl nodes in FM 2 seen in the polar plots {(Figs. 4g-4j)} is striking. The figure-eights of $\sigma_{xx}$ and $\sigma_{yx}$ are now elongated along $\varphi = 30^\circ$ direction at high fields,
corresponding to a larger tilt {(Fig. 4f)} and further rotation of \textbf{\textit{\^{t}}} in the $xy$-plane.
The $\varphi$-variations of $\sigma_{xx}$ and $\sigma_{yx}$  at low fields differ drastically  --- while the long axis of $\sigma^{avg}_{xx}(\varphi)$  points in the $\varphi = 60^\circ$ direction, corresponding (as in FM 1) to a higher $\alpha$-value at low fields, the $\sigma_{yx}(\varphi)$ has radically shape-changed to feature antisymmetry {(Fig. 4j)} and an order-of-magnitude lower value of $\alpha$.
{Concurrently, the in-plane transport displays strong enhancements of field-rotational chirality $\chi^{\pm \varphi}$ {(Fig. 4m,n and Extended Data Figs. 8,9)},
consistent with the low-dissipation $\varphi$-chiral transport amplified at low fields.}
{And, while theory fully describing the newly created Weyl phases is yet to be developed we infer that low $\alpha$, which is stabilized on the subsequent hydrogen cycling ({Fig. S12}),
{attests to the chirality-enhanced} origins of {linear-in-current} {(Fig. S13)} field-antisymmetric magnetoresistance in the hydrogen-altered MST.

{Lastly, we remark that our work represents an advance in designer topological quantum materials.
Topological transformations facilitated by hydrogen or other light elements (such as e.g., lithium \cite{Lithium-1990}) through defect-related pathways
expand the availability of robust and easily accessible platforms for harnessing distinct topological phases with stunning macroscopic behaviors, thereby opening
a path to a potentially disruptive chirality-based practical implementations}.

\vspace{10mm}
\def\baselinestretch{1.35}

\newpage
\def\baselinestretch{1.36}
\begin{center}
\section*{FIGURE LEGENDS}
\end{center}
\vspace{-2mm}
\normalsize
\noindent \textbf{Figure 1. $\mid$ Hydrogen-induced transitions between disparate spin orders in MnSb$_2$Te$_4$.}
\textbf{a}, Longitudinal resistance $R_{xx}$ and magnetization $M$ \textit{vs.} temperature of as-grown ferromagnetic MnSb$_2$Te$_4$ (MST). The cusp in $R_{xx}(T)$ at the Curie temperature $T_C \cong 26~\textrm{K}$
coincides with the maximum slope in $M(T)$.  The field is out-of-plane ($\parallel{c}$-axis). \textit{Inset:} Optical image of the van der Pauw contact geometry used.
\textbf{b}, Illustration of hydrogenation: sample is submerged in a dilute aqueous HCl solution (0.5M) at room temperature where H$^+$ permeates the bulk through a timed diffusion process.
\textbf{c}, Hydrogenation turns the as-grown MST ({FM~1}) into a robust antiferromagnet (AFM) {(also  Extended Data Fig. 4)} with N\'{e}el temperature $T_N \approx 20~\textrm{K}$, akin to as-grown AFM MST.
A typical AFM behavior is evident in $R_{xx}(T)$ and $M(T)$. {$M$ was measured on thicker samples while transport was measured on exfoliated flakes (see Methods}).
\textbf{d}, Transverse (Hall) signal $R_{yx}(T)$  tracks AFM magnetization in (\textbf{c}). \textit{Inset:} {A characteristic AFM quasi-plateau \cite{Otrokov2019} in $R_{yx}(H)$ after hydrogenation. }
\textbf{e}, Illustration of dehydrogenation: H$^+$ inside the sample is released as H$_2$ gas by an anneal in the $100-130^\circ\textrm{C}$ temperature range prescribed by the sample thickness.
\textbf{f}, Temperature evolution of the Hall resistance $R_{yx}$ after release of hydrogen. Dehydrogenation returns the system to a ferromagnetic state (FM~2) with  a nearly doubled Curie temperature $T_C$.
\textit{Inset:} $R_{yx}$ hysteresis loop in as-grown MST.
\textbf{g}, A typical \textit{H-symmetric} longitudinal resistance $R_{xx}$ with two peaks (\textit{p-p}) at the magnetization reversal in the FM~1 state. { \textbf{h}, Cartoons of average magnetic alignments in FM 1 and  FM 2. }
\textbf{i}, $R_{xx}(H)$ in {FM~2} is \textit{H-antisymmetric}; it has flipped into a dip and a peak (\textit{d-p}).

\vspace{3mm}
\normalsize
\noindent \textbf{Figure ~2 $\mid$ Hydrogen-induced transport anisotropy of ferromagnetic MnSb$_2$Te$_4$ under tilted magnetic fields}.
\textbf{a,b}, Longitudinal conductance $\sigma_{xx}$ under the field tilt
in the {FM~1} and {FM~2} states respectively. Polar plots of $\sigma_{xx}$ show magnetoconductance favoring the \textit{c}-axis in {FM~1},
while in {FM~2} the preferred axis of $45^\circ$ appears at low fields.
\textbf{c,d}, Illustration of the relative low-field alignments of magnetic field $H$, magnetization $M$, and the average direction of domain walls in the {FM~1} and {FM~2} states respectively.
\textbf{e,f}, Low-field 3D contour plots of $R_{xx}$ and $R_{yx}$ in the $H-\theta$ plane for the {FM~1} state.
\textbf{g,h}, Low-field 3D contour plots of $R_{xx}$ and $R_{yx}$ for the {FM~2} state.
In {FM~1} the resistance is maximum when the field is in-plane ($\theta = 90^\circ$). In {FM~2} a `negative'
$R_{xx}$ signal is detected at $\theta = 45^\circ$, and a `positive' signal follows (dip and peak).
The response now is antisymmetric across $180^\circ$.
\textbf{h}, The low-field asymmetry of {FM~2} shows up in the Hall signal $R_{yx}$ in the $H-\theta$ plane.
The in-plane transverse (Planar Hall) signal $R_{yx}(H)$ --- in \textbf{i}, for {FM~1} and \textbf{j}, for {FM~2} (see text).
A zero-field-centered hysteresis loop with a width $H^{ab}_c$ is
tiny in {FM~1} but becomes large in {FM~2}.
\textbf{k}, Temperature evolution of %the planar hysteresis loop widths
$H^{ab}_c$.  In {FM~2}, $H^{ab}_c$ nearly vanishes at $T_{dp} \sim 40~\textrm{K}$, but  then reenters  at the level of {FM~1} to vanish at $T_C$.
{\textbf{l}, $\Delta^{p-d}R_{xx}$, the resistance difference between the dip and the peak, disappears at $T_{dp}$ at which the $H$-symmetric  $R_{xx}$ is recovered. }
\textbf{m,n}, A direct connection between $H$-antisymmetric $R_{xx}$ and the PHE signal $R_{yx}(H)$. \textbf{m}, Evolution of the peak and dip (\textit{p-d}) fields with the tilt angle $\theta$ at 2 K. %It shows their widening separation.
\textbf{n}, For $H\parallel ab$, \textit{d-p} transfigures into the `wings' of  the symmetrized PHE.

\vspace{3mm}
\noindent \textbf{Figure~3 $\mid$ In-plane charge transport in the as-grown and hydrogen-modified FM states of MnSb$_2$Te$_4$}.
\textbf{a}, Field dependence of the longitudinal conductance $\sigma_{xx}$ in as-grown ferromagnetic MnSb$_2$Te$_4$ (FM~1) at 2 K.
Beyond 200 mT, $\sigma_{xx}(H)$) is {nearly} \textit{quadratic} in $H$. \textit{Inset:} In the magnetization reversal region at low fields ($H < 200~ \textrm{mT}$) $\sigma_{xx}(H)$ is quasilinear.
\textbf{b}, {Field dependence of $\sigma_{xx}$ in hydrogen-modified ferromagnetic MnSb$_2$Te$_4$ (FM~2) at 2 K. Beyond 200 mT, $\sigma_{xx}(H)$) is \textit{linear} in $H$. \textit{Inset:}
The  antisymmetric MC at low fields now comprises the center hysteresis loop and a dip and peak features originating from the PHE signal shown in Fig. 2j.
\textbf{c}, 3D plot of the PHE conductance $\sigma_{yx}(H,\varphi)$ in FM 1 over a large field and angle range.
\textbf{d}, 3D plot of the PHE conductance $\sigma_{yx}(H,\varphi)$ in FM 2 over a large field and angle range shows it to be field-antisymmetric about $H = 0$ (see text).
\textbf{e}, In-plane angular ($\varphi$) dependence of transport in FM 1 at 9 T obtained by rotating field in plane (see cartoon).
Top: Planar Hall effect (PHE) $\sigma_{yx}(\varphi) \propto \sin\varphi$. Bottom: $\Delta \sigma^{avg}_{xx}(\varphi) \propto \cos^2\varphi$, the $\varphi$-change in field-averaged longitudinal magnetoconductance (MC)
$\sigma^{avg}_{xx} = \frac{1}{2}(\sigma_{xx} (H) +  \sigma_{xx} (-H))$  with the background subtracted for each value of $H$.
\textbf{f}, The planar Hall signal (PHE) in FM 1 is large; the ratio of PHE to the change in MC reaches 1000$\%$ at low fields.
\textbf{g}, The in-plane angular dependencies of $\sigma_{yx}(\varphi)$ (top) and $\Delta\sigma^{avg}_{xx}(\varphi)$ (bottom) in FM 2. Although they are similar to those
in FM 1,  the deviations are clearly evident.
\textbf{h}, PHE signal in FM 2 is further enhanced by an order of magnitude.
\textbf{i}, The two frequencies in the FFT of $\sigma_{yx}(\varphi)$ in FM 2 indicate that during the $\varphi$-rotation another angle corresponding to a new symmetry axis enters.}

\vspace{3mm}
\noindent \textbf{Figure~4 $\mid$ Symmetry change and Weyl cone tilt revealed under in-plane field rotation in hydrogen-modified MnSb$_2$Te$_4$}.
{\textbf{a}, Illustration of energy-momentum dispersion for a pair of oppositely tilted Type I Weyl nodes representing FM~1.
The position of the Fermi level in our \textit{p}-type material is indicated as grey slab.
Polar plots of the in-plane angular dependencies of $\sigma^{avg}_{xx}(\varphi)$ in FM~1 \textbf{b}, at 4 T , and \textbf{c}, at 0.5 T.
{The slanted `figure-eight' at 0.5 T is accounted for by an increase in the internode/intranode
scattering strength $\alpha$, {shown in the outsets of the corresponding theory plots} {(Supplementary Section II).}}
Polar plots of $\sigma_{yx}(\varphi)$  in FM~1 \textbf{d}, at 4 T, and \textbf{e}, at 0.5 T, for positive (red) and negative (blue) fields.
The values of $\alpha$  are in correspondence with $\sigma_{xx}(\varphi)$.
\textbf{f}, An energy-momentum dispersion for oppositely tilted and rotated Weyl nodes representing FM~2.
In-plane polar plots of $\sigma^{avg}_{xx}(\varphi)$ in FM~2 \textbf{g}, at 4 T  and \textbf{h}, at 0.5 T.
In-plane polar plots of $\sigma_{yx}(\varphi)$ in FM~2 for positive and negative fields: \textbf{i}, at 4 T and \textbf{j}, at 0.5 T.
{\textbf{k,l}}, Cartoons of three directed quantities: electric field \textit{{\^{E}}}, magnetic field \textit{{\^{H}}},
and the Weyl cone tilt direction \textit{{\^{R}}} with its projection \textbf{\textit{\^{t}}}  in the $xy$-plane (see text). {Tilt values obtained from theory are indicated.}
$\gamma$ is the angle between tilt \textbf{\textit{\^{t}}} and magnetic field.
{A much larger and rotated tilt \textbf{\textit{\^{t}}} is represented in FM~2. }
{\textbf{m,n}, $\varphi$ chirality $\chi^{\pm \varphi}$ of planar Hall signal $\sigma_{yx}(\varphi)$  and  the width of the planar hysteresis loop $H_c$ in FM 1 and FM 2.
Strong chirality amplification is seen in FM 2 outside of the characteristic angle ($45-60^\circ$) range, {see Extended Data Figs. 8,9}.  {For a subsequent hydrogenation cycle see {Fig. S12}}.}

\newpage

\vspace{5mm}
\hspace{-10mm}
\includegraphics[width=1.0\textwidth]{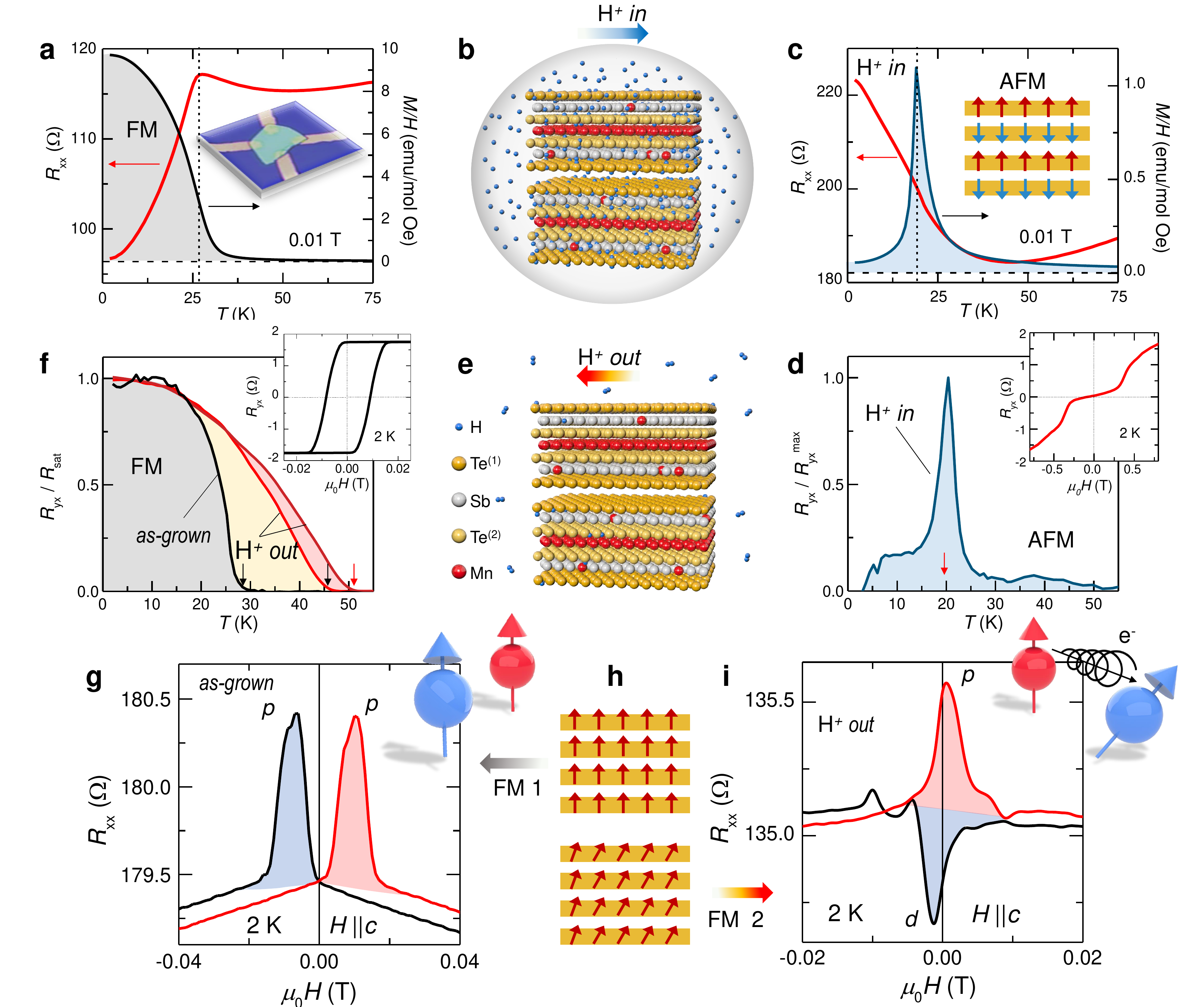}
\vfill\hfill Fig.~1; {AT \it et al.} \eject

\vspace{25mm}
\hspace{-20mm}
\includegraphics[width=1.15\textwidth]{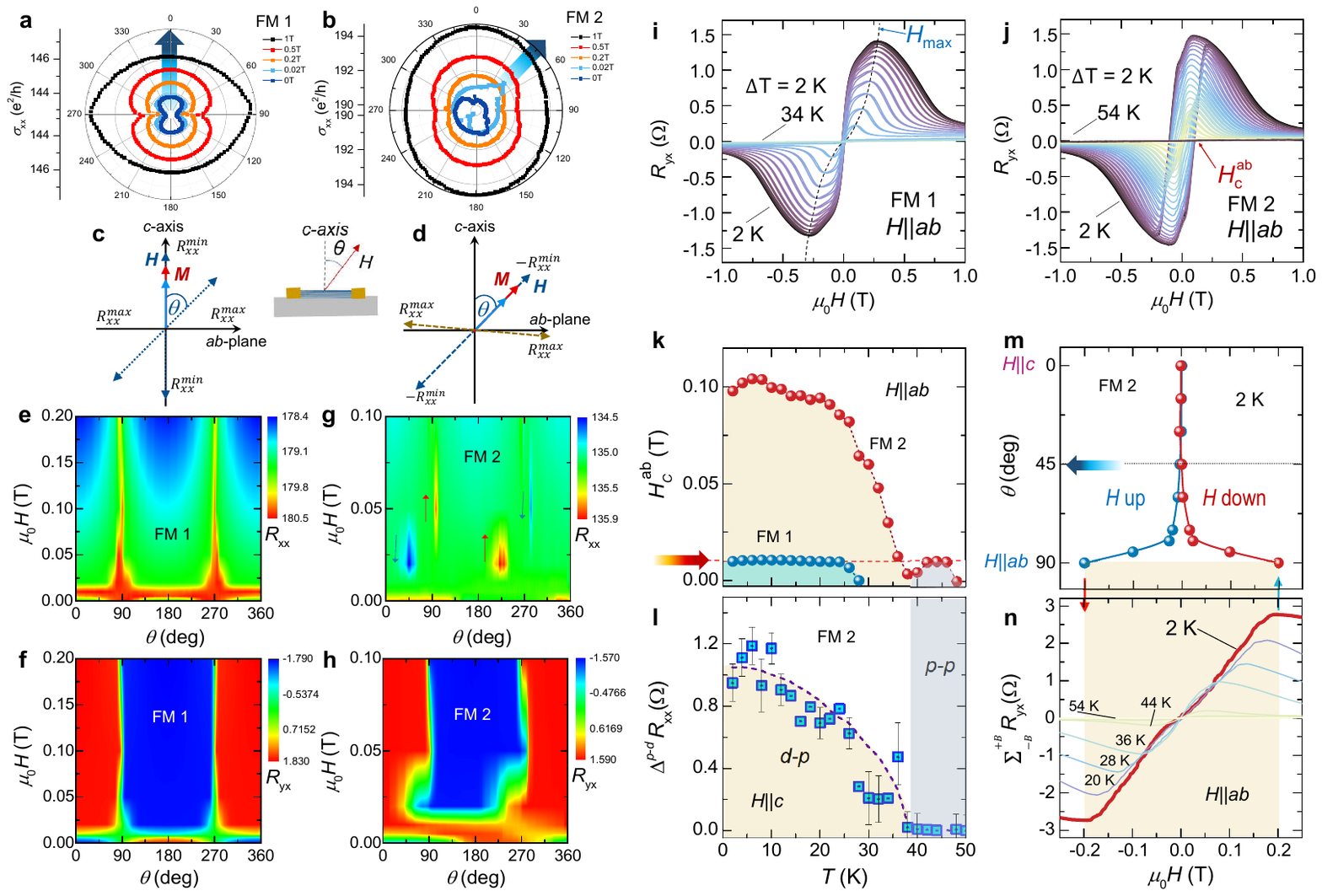}
\vfill\hfill Fig.~2; {AT \it et al.} \eject

\vspace{25mm}
\hspace{-5mm}
\includegraphics[width=0.9\textwidth]{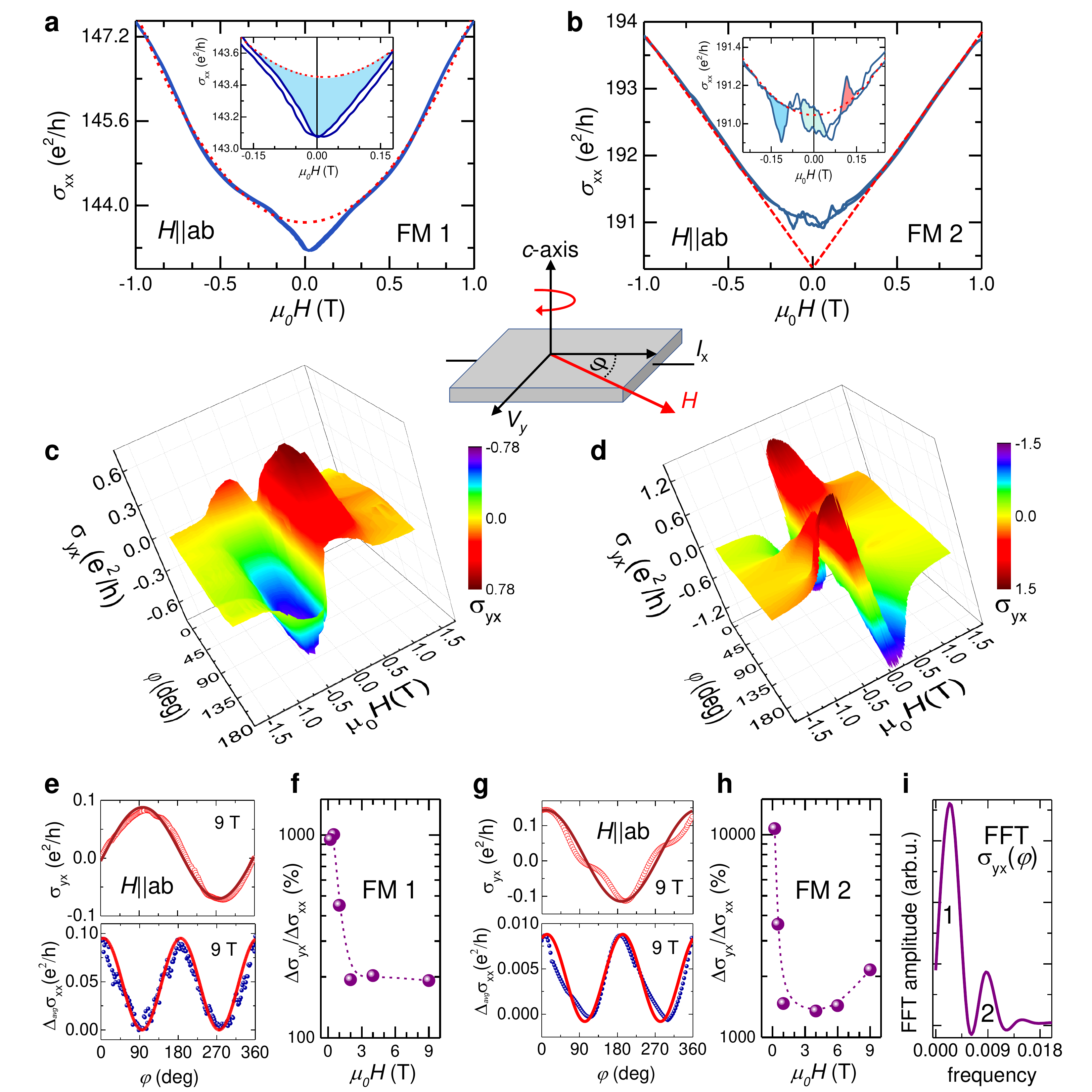}
\vfill\hfill Fig.~3; {AT \it et al.} \eject

\vspace{25mm}
\hspace{-11mm}
\includegraphics[width=1.0\textwidth]{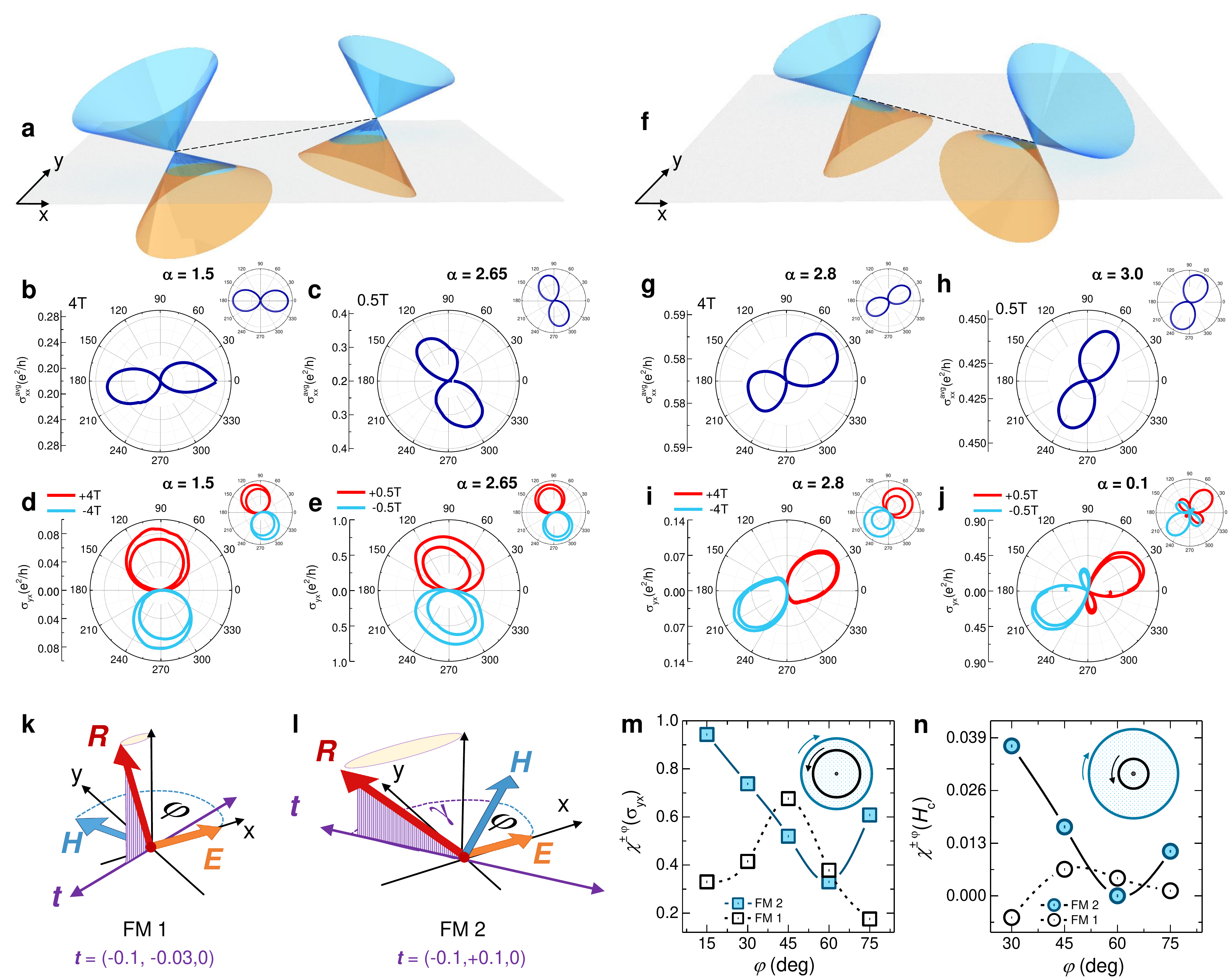}
\vfill\hfill Fig.~4; {AT \it et al.} \eject

\begin{center}
\section*{REFERENCES}
\end{center}
\vspace{-12mm}
%\begin{thebibliography}{10}


\begin{thebibliography}{99}
\expandafter\ifx\csname url\endcsname\relax
  \def\url#1{\texttt{#1}}\fi
\expandafter\ifx\csname urlprefix\endcsname\relax\def\urlprefix{URL }\fi
\providecommand{\bibinfo}[2]{#2}
\providecommand{\eprint}[2][]{\url{#2}}

\bibitem{Chiral2-1990} %1
\bibinfo{author}{Kondepudi, D.} \& \bibinfo{author}{Hegstrom, R.}
\newblock \bibinfo{title}{{The handedness of the universe.}}
\newblock \emph{\bibinfo{journal}{Scientific American}}
  \textbf{\bibinfo{volume}{262}}, \bibinfo{pages}{108--115}
  (\bibinfo{year}{1990}).

\bibitem{chiral-bio2001} %2
\bibinfo{author}{Siegel, J.}
\newblock \bibinfo{title}{{Single-handed cooperation.}}
\newblock \emph{\bibinfo{journal}{Nature}} \textbf{\bibinfo{volume}{409}},
  \bibinfo{pages}{777--778} (\bibinfo{year}{2001}).

\bibitem{chiral-antimatter2003}%3
\bibinfo{author}{Ellis, J.}
\newblock \bibinfo{title}{{Antimatter matters.}}
\newblock \emph{\bibinfo{journal}{Nature}} \textbf{\bibinfo{volume}{424}},
  \bibinfo{pages}{631--634} (\bibinfo{year}{2003}).

\bibitem{Weyl-review2018} %4
\bibinfo{author}{Armitage, N.}, \bibinfo{author}{Mele, E.} \&
  \bibinfo{author}{Vishwanath, A.}
\newblock \bibinfo{title}{{Weyl and Dirac semimetals in three-dimensional
  solids.}}
\newblock \emph{\bibinfo{journal}{Rev. Mod. Phys.}}
  \textbf{\bibinfo{volume}{90}}, \bibinfo{pages}{187001}
  (\bibinfo{year}{2018}).

\bibitem{Chir-DWlogic-Omari2019} %5
\bibinfo{author}{Omari, K.} \emph{et~al.}
\newblock \bibinfo{title}{{Toward chirality-encoded domain wall logic.}}
\newblock \emph{\bibinfo{journal}{Adv. Funct. Mater.}}
  \textbf{\bibinfo{volume}{29}}, \bibinfo{pages}{1807282}
  (\bibinfo{year}{2019}).

\bibitem{Chir-spin_Parkin2021} %6
\bibinfo{author}{Yang, S.-H.}, \bibinfo{author}{Naaman, R.},
  \bibinfo{author}{Paltiel, Y.} \& \bibinfo{author}{Parkin, S.}
\newblock \bibinfo{title}{{Chiral spintronics.}}
\newblock \emph{\bibinfo{journal}{Nature Rev. Phys.}}
  \textbf{\bibinfo{volume}{3}}, \bibinfo{pages}{328–343}
  (\bibinfo{year}{2021}).

\bibitem{BerryPhaseReview2010} %7
\bibinfo{author}{Xiao, D.}, \bibinfo{author}{Chang, M.-C.} \&
  \bibinfo{author}{Niu, Q.}
\newblock \bibinfo{title}{{Berry phase effects on electronic properties.}}
\newblock \emph{\bibinfo{journal}{Rev. Mod. Phys.}}
  \textbf{\bibinfo{volume}{82}}, \bibinfo{pages}{1959–2007}
  (\bibinfo{year}{2010}).

  \bibitem{PHE-Burkov2017} %8
\bibinfo{author}{Burkov, A.}
\newblock \bibinfo{title}{{Giant planar Hall effect in topological metals.}}
\newblock \emph{\bibinfo{journal}{Phys. Rev. B}} \textbf{\bibinfo{volume}{95}},
  \bibinfo{pages}{041110(R)} (\bibinfo{year}{2017}).

\bibitem{Weyl1929}%9
\bibinfo{author}{Weyl, H.}
\newblock \bibinfo{title}{{Elektron und Gravitation. I.}}
\newblock \emph{\bibinfo{journal}{Z. Physik}} \textbf{\bibinfo{volume}{56}},
  \bibinfo{pages}{330–352} (\bibinfo{year}{1929}).

\bibitem{Ninomiya1983} %10
\bibinfo{author}{Nielsen, H.} \& \bibinfo{author}{Ninomiya, M.}
\newblock \bibinfo{title}{{The Adler-Bell-Jackiw anomaly and Weyl fermions in a
  crystal.}}
\newblock \emph{\bibinfo{journal}{Phys. Lett. B}}
  \textbf{\bibinfo{volume}{130}}, \bibinfo{pages}{389} (\bibinfo{year}{1983}).

\bibitem{Hasan2015}%11
\bibinfo{author}{Xu, S.-Y.} \emph{et~al.}
\newblock \bibinfo{title}{{Discovery of a Weyl fermion semimetal and
  topological Fermi arcs.}}
\newblock \emph{\bibinfo{journal}{Science}} \textbf{\bibinfo{volume}{349}},
  \bibinfo{pages}{613–617} (\bibinfo{year}{2015}).

\bibitem{Weyl-review2021}%12
\bibinfo{author}{Hasan, M.} \emph{et~al.}
\newblock \bibinfo{title}{{Weyl, Dirac and high-fold chiral fermions in
  topological quantum matter.}}
\newblock \emph{\bibinfo{journal}{Nature Rev. Phys.}}
  \textbf{\bibinfo{volume}{6}}, \bibinfo{pages}{784–803}
  (\bibinfo{year}{2021}).

\bibitem{WEYL-TypeII2015}
\bibinfo{author}{Soluyanov, A.} \emph{et~al.}
\newblock \bibinfo{title}{{Type-II Weyl semimetals.}}
\newblock \emph{\bibinfo{journal}{Nature}} \textbf{\bibinfo{volume}{527}},
  \bibinfo{pages}{495–498} (\bibinfo{year}{2015}).

\bibitem{Weyl-TypeII2016}
\bibinfo{author}{Huang, L.} \emph{et~al.}
\newblock \bibinfo{title}{{Spectroscopic evidence for a Type II Weyl
  semimetallic state in MoTe$_2$.}}
\newblock \emph{\bibinfo{journal}{Nature Mater.}}
  \textbf{\bibinfo{volume}{15}}, \bibinfo{pages}{1155–1160}
  (\bibinfo{year}{2016}).

\bibitem{hybrid-Weyl2016}
\bibinfo{author}{Li, F.-Y.} \emph{et~al.}
\newblock \bibinfo{title}{{Hybrid Weyl semimetal.}}
\newblock \emph{\bibinfo{journal}{Phys. Rev. B}} \textbf{\bibinfo{volume}{94}},
  \bibinfo{pages}{121105(R)} (\bibinfo{year}{2016}).

\bibitem{higher-orderWeyl2020}
\bibinfo{author}{Ghorashi, S.}, \bibinfo{author}{Li, T.} \&
  \bibinfo{author}{Hughes, T.}
\newblock \bibinfo{title}{{Higher-order Weyl semimetals.}}
\newblock \emph{\bibinfo{journal}{Phys. Rev. Lett.}}
  \textbf{\bibinfo{volume}{125}}, \bibinfo{pages}{266804}
  (\bibinfo{year}{2020}).

\bibitem{mKagome2019}
\bibinfo{author}{Liu, D.} \emph{et~al.}
\newblock \bibinfo{title}{{Magnetic Weyl semimetal phase in a Kagom$\acute{e}$
  crystal.}}
\newblock \emph{\bibinfo{journal}{Science}} \textbf{\bibinfo{volume}{365}},
  \bibinfo{pages}{1282–1285} (\bibinfo{year}{2019}).

\bibitem{FermiArcs2019}
\bibinfo{author}{Morali, N.} \emph{et~al.}
\newblock \bibinfo{title}{{Fermi-arc diversity on surface terminations of the
  magnetic Weyl semimetal Co$_3$Sn$_2$S$_2$.}}
\newblock \emph{\bibinfo{journal}{Science}} \textbf{\bibinfo{volume}{365}},
  \bibinfo{pages}{1286–1291} (\bibinfo{year}{2019}).

\bibitem{TypeII-Bernevig2016}
\bibinfo{author}{Bradlyn, B.} \emph{et~al.}
\newblock \bibinfo{title}{{Beyond Dirac and Weyl fermions: Unconventional
  quasiparticles in conventional crystals.}}
\newblock \emph{\bibinfo{journal}{Science}} \textbf{\bibinfo{volume}{353}},
  \bibinfo{pages}{aaf5037} (\bibinfo{year}{2016}).

\bibitem{axial2020}
\bibinfo{author}{Ilan, R.}, \bibinfo{author}{Grushin, A.} \&
  \bibinfo{author}{Pikulin, D.}
\newblock \bibinfo{title}{{Pseudo-electromagnetic fields in 3D topological
  semimetals.}}
\newblock \emph{\bibinfo{journal}{Nature Rev. Phys.}}
  \textbf{\bibinfo{volume}{2}}, \bibinfo{pages}{29–41}
  (\bibinfo{year}{2020}).

\bibitem{mWeylsAraki2020}
\bibinfo{author}{Araki, Y.}
\newblock \bibinfo{title}{{Magnetic textures and dynamics in magnetic Weyl
  semimetals.}}
\newblock \emph{\bibinfo{journal}{Annalen der Phys.}}
  \textbf{\bibinfo{volume}{532}}, \bibinfo{pages}{1900287}
  (\bibinfo{year}{2020}).

\bibitem{HCl-Haiming2022}
\bibinfo{author}{Deng, H.} \emph{et~al.}
\newblock \bibinfo{title}{{Topological surface currents accessed through
  reversible hydrogenation of the three-dimensional bulk.}}
\newblock \emph{\bibinfo{journal}{Nature Comms.}}
  \textbf{\bibinfo{volume}{13}}, \bibinfo{pages}{2308} (\bibinfo{year}{2022}).

\bibitem{tilted_Weyls2019}
\bibinfo{author}{Das, K.} \& \bibinfo{author}{Agarwal, A.}
\newblock \bibinfo{title}{{Linear magnetochiral transport in tilted type-I and
  type-II Weyl semimetals.}}
\newblock \emph{\bibinfo{journal}{Phys. Rev. B}} \textbf{\bibinfo{volume}{99}},
  \bibinfo{pages}{085405} (\bibinfo{year}{2019}).

\bibitem{AntisymMR-Weyl2021}
\bibinfo{author}{Jiang, B.} \emph{et~al.}
\newblock \bibinfo{title}{{Chirality-dependent Hall effect and antisymmetric
  magnetoresistance in a magnetic Weyl semimetal.}}
\newblock \emph{\bibinfo{journal}{Phys. Rev. Lett.}}
  \textbf{\bibinfo{volume}{126}}, \bibinfo{pages}{236601}
  (\bibinfo{year}{2021}).

\bibitem{Berry-antisym2022}
\bibinfo{author}{Zeng, Q.} \emph{et~al.}
\newblock \bibinfo{title}{{Berry curvature induced antisymmetric in-plane
  magneto-transport in magnetic Weyl EuB$_6$.}}
\newblock \emph{\bibinfo{journal}{Appl. Phys. Lett.}}
  \textbf{\bibinfo{volume}{121}}, \bibinfo{pages}{162405}
  (\bibinfo{year}{2022}).

\bibitem{MBT-family2019}
\bibinfo{author}{Li, J.} \emph{et~al.}
\newblock \bibinfo{title}{{Intrinsic magnetic topological insulators in van der
  Waals layered MnBi$_2$Te$_4$-family materials.}}
\newblock \emph{\bibinfo{journal}{Sci. Adv.}} \textbf{\bibinfo{volume}{5}},
  \bibinfo{pages}{eaaw5685} (\bibinfo{year}{2019}).

\bibitem{axionMBT2020}
\bibinfo{author}{Liu, C.} \emph{et~al.}
\newblock \bibinfo{title}{{Robust axion insulator and Chern insulator phases in
  a two-dimensional antiferromagnetic topological insulator.}}
\newblock \emph{\bibinfo{journal}{Nature Mater.}}
  \textbf{\bibinfo{volume}{19}}, \bibinfo{pages}{522–527}
  (\bibinfo{year}{2020}).

\bibitem{QAH-Haiming2020}
\bibinfo{author}{Deng, H.} \emph{et~al.}
\newblock \bibinfo{title}{{High-temperature quantum anomalous Hall regime in a
  MnBi$_2$Te$_4$/Bi$_2$Te$_3$ superlattice.}}
\newblock \emph{\bibinfo{journal}{Nature Phys.}} \textbf{\bibinfo{volume}{17}},
  \bibinfo{pages}{36–42} (\bibinfo{year}{2021}).

\bibitem{QAH-MBT-Science2020}
\bibinfo{author}{Deng, Y.} \emph{et~al.}
\newblock \bibinfo{title}{{Quantum anomalous Hall effect in intrinsic magnetic
  topological insulator MnBi$_2$Te$_4$.}}
\newblock \emph{\bibinfo{journal}{Science}} \textbf{\bibinfo{volume}{367}},
  \bibinfo{pages}{895–900} (\bibinfo{year}{2020}).

\bibitem{MBST-phase_diagram2019}
\bibinfo{author}{Chen, B.} \emph{et~al.}
\newblock \bibinfo{title}{{Intrinsic magnetic topological insulator phases in
  the Sb doped MnBi$_2$Te$_4$ bulks and thin flakes.}}
\newblock \emph{\bibinfo{journal}{Nature Comms.}}
  \textbf{\bibinfo{volume}{10}}, \bibinfo{pages}{4469} (\bibinfo{year}{2019}).

\bibitem{AFM-MST-typeIIWeyl2021}
\bibinfo{author}{Lee, S.} \emph{et~al.}
\newblock \bibinfo{title}{{Evidence for a magnetic-field-induced ideal Type-II
  Weyl state in antiferromagnetic topological insulator
  Mn(Bi$_{1-x}$Sb$_x$)$_2$Te$_4$.}}
\newblock \emph{\bibinfo{journal}{Phys. Rev. X}} \textbf{\bibinfo{volume}{11}},
  \bibinfo{pages}{031032} (\bibinfo{year}{2021}).

\bibitem{MST-Weyl_Murakami2019}
\bibinfo{author}{Murakami, T.} \emph{et~al.}
\newblock \bibinfo{title}{{Realization of interlayer ferromagnetic interaction
  in MnSb$_2$Te$_4$ toward the magnetic Weyl semimetal state.}}
\newblock \emph{\bibinfo{journal}{Phys. Rev. B}}
  \textbf{\bibinfo{volume}{100}}, \bibinfo{pages}{195103}
  (\bibinfo{year}{2019}).

\bibitem{Otrokov2019}
\bibinfo{author}{Otrokov, M.} \emph{et~al.}
\newblock \bibinfo{title}{{Prediction and observation of an antiferromagnetic
  topological insulator.}}
\newblock \emph{\bibinfo{journal}{Nature}} \textbf{\bibinfo{volume}{576}},
  \bibinfo{pages}{416–422} (\bibinfo{year}{2019}).

\bibitem{MST-site_mixing-Liu2021}
\bibinfo{author}{Liu, Y.} \emph{et~al.}
\newblock \bibinfo{title}{{Site mixing for engineering magnetic topological
  insulators.}}
\newblock \emph{\bibinfo{journal}{Phys. Rev. X}} \textbf{\bibinfo{volume}{11}},
  \bibinfo{pages}{021033} (\bibinfo{year}{2021}).

\bibitem{Wimmer2021}
\bibinfo{author}{Wimmer, S.} \emph{et~al.}
\newblock \bibinfo{title}{{Mn–Rich MnSb$_2$Te$_4$: A topological insulator
  with magnetic gap closing at high Curie temperatures of 45–50 K.}}
\newblock \emph{\bibinfo{journal}{Adv. Mater.}} \textbf{\bibinfo{volume}{33}},
  \bibinfo{pages}{2102935} (\bibinfo{year}{2021}).

\bibitem{in-plane-AHE-Nature2022}
\bibinfo{author}{Zhou, J.} \emph{et~al.}
\newblock \bibinfo{title}{{Heterodimensional superlattice with in-plane
  anomalous Hall effect.}}
\newblock \emph{\bibinfo{journal}{Nature}} \textbf{\bibinfo{volume}{609}},
  \bibinfo{pages}{46–51} (\bibinfo{year}{2022}).

\bibitem{Tewari_PRL2017}
\bibinfo{author}{Nandy, S.}, \bibinfo{author}{Sharma, G.},
  \bibinfo{author}{Taraphder, A.} \& \bibinfo{author}{Tewari, S.}
\newblock \bibinfo{title}{{Chiral anomaly as the origin of the planar Hall
  effect in Weyl semimetals.}}
\newblock \emph{\bibinfo{journal}{Phys. Rev. Lett.}}
  \textbf{\bibinfo{volume}{119}}, \bibinfo{pages}{176804}
  (\bibinfo{year}{2017}).

\bibitem{Tewari_PRB2023}
\bibinfo{author}{Sharma, G.}, \bibinfo{author}{Nandy, S.},
  \bibinfo{author}{Raman, K.} \& \bibinfo{author}{Tewari, S.}
\newblock \bibinfo{title}{{Decoupling intranode and internode scattering in
  Weyl fermions.}}
\newblock \emph{\bibinfo{journal}{Phys. Rev. B}}
  \textbf{\bibinfo{volume}{107}}, \bibinfo{pages}{115161}
  (\bibinfo{year}{2023}).

\bibitem{Lithium-1990}
\bibinfo{author}{Myakenkaya, G.}, \bibinfo{author}{Gutsev, G.},
  \bibinfo{author}{Afanaseva, N.}, \bibinfo{author}{Evseev, V.} \&
  \bibinfo{author}{Konopleva, R.}
\newblock \bibinfo{title}{{Lithium interaction with lattice defects in Si.}}
\newblock \emph{\bibinfo{journal}{Phys. Stat. Sol.(b)}}
  \textbf{\bibinfo{volume}{161}}, \bibinfo{pages}{91–103}
  (\bibinfo{year}{1990}).

\clearpage


\end{thebibliography}
\end{document}